\documentclass[runningheads,a4paper]{llncs}

\usepackage{amsmath}
\usepackage{amssymb}
\usepackage{graphicx}
\usepackage{placeins} 
\usepackage{url}

\newtheorem{constraint}{Constraint}

\urldef{\mailsa}\path|william.stevens@uwe.ac.uk|
\newcommand{\keywords}[1]{\par\addvspace\baselineskip
\noindent\keywordname\enspace\ignorespaces#1}
\begin{document}

\mainmatter

\title{Using transition systems to describe and predict the behaviour of structured excitable media}
\titlerunning{Transition systems for excitable media}
\author{William M Stevens}
\institute{Unconventional Computing Group, Faculty of Environment and Technology, University of the West of England, Frenchay Campus, Coldharbour Lane, Bristol, BS16 1QY\\
\mailsa\\
}
\toctitle{Using transition systems to describe and predict the behaviour of structured excitable media}
\tocauthor{William M Stevens}

\maketitle              % typeset the title of the contribution
\begin{abstract}
I show how transition systems can be applied to the naturally concurrent behaviour of excitable media. I consider structured excitable media, in which excitations are constrained to propagate only in defined narrow channels, and cannot propagate elsewhere. I define a type of transition system that can be used to describe the complete set of behaviours exhibited by simple structures. The composition rules that result from this definition can be used to automatically deduce the behaviour of more complex structures composed from simpler structures. Several examples illustrate the method, and a software implementation is provided.
\keywords{Transition systems, Concurrency, Excitable media, Toppling dominoes, Physarum polycephalum, Belousov-Zhabotinsky reaction}
\end{abstract}
\section{Introduction}

Unconventional computing seeks to explore and characterise a wider range of substrates and phenomena than those that have been used so far for practical computing, with the hope of understanding and applying the intrinsic information processing behaviour of each substrate. Unconventional computing tends to be substrate-driven rather than model-driven. Rather than searching for physical systems that can be harnessed to make computing machinery according to a preconceived idea of which behaviour the machinery is required to exhibit, unconventional computing is inspired by the idea that every sufficiently complex system performs information processing of some kind, which we can harness if we can find the appropriate mathematical techniques for understanding or interpreting the system, and the appropriate experimental techniques for manipulating and observing the system. The existence of natural systems that manage to harness complex phenomena to process information in powerful ways that we do not yet fully understand is one of the motivations for this search, though unconventional computing does not confine itself to naturally occurring systems.

One of the difficulties in efficiently exploiting previously unexploited phenomena in new or existing substrates is that many of the abstract models that we use in computer science were invented and popularised precisely because they abstract away from the details of the physical behaviour of computing hardware. In the early history of computing machinery it was often pragmatic to regard concurrent, time-dependent behaviour as a nuisance to be abstracted away from so that it did not complicate and impede design. In synchronous circuit design the input-output delay present in all logic gates is taken account of by waiting for a sufficiently long time that all gates have had enough time to propagate any changes on their inputs to their outputs, and only then carrying out the next step in the computation. Asynchronous design techniques \cite{sparso2001} can help designers to make more efficient designs by having circuits explicitly signal when their outputs become valid, so that there is no need to wait for the worst-case delay time, and no need to have all circuits in a system clocked at constant rates.

Asynchronous circuit design involves understanding electronic digital circuit elements at a less abstract level than that which is used in synchronous design, this makes it possible to exploit types of behaviour that do not exist at higher levels of abstraction. This approach towards the substrate can be generalised \cite{stepney2008}. Some of the questions that can be asked of any substrate considered to have complex information processing potential are:

\begin{enumerate}
\item What are the intrinsic behaviours of a region of substrate and how can they be modelled?
\item How do regions of substrate behave when connected together?
\item Can the modes of behaviour of regions of substrate, or composed regions of substrate be mapped easily onto any existing abstractions used for information processing?
\item Does the substrate suggest new models for information processing?
\item Can the substrate be applied to a problem without having to have a detailed abstract model for understanding how it behaves as it does?
\end{enumerate}

For complex substrates answers to questions 3,4 and 5 can be sought through observation and experimentation without having quantitative answers for questions 1 and 2. For example, the tendency of plasmodium of physarum polycephalum to form efficient networks connecting multiple food sources was justified on evolutionary grounds and observed experimentally \cite{nakagaki2004} before a mathematical model of dynamic network formation was constructed \cite{tero2007}, and before a low-level particle based model that can give rise to that behaviour was constructed \cite{jones2010}. Robots have been devised that have controllers which exploit a substrate without requiring a theory that accounts for the behaviour of the substrate \cite{tsuda2007,harding2005}.

For other substrates, the behaviour of small regions of substrate may be easy to explain and describe \cite{stevens2008}, but challenges arise with recognising that the substrate might be capable of information processing, with mapping it onto higher level abstractions, and with trying to deduce the behaviour of large regions of substrate from the behaviour of small regions.

In this paper I consider excitable media substrates, which can be used to implement channels along which excitations can propagate. Examples of this type of substrate are: planar Belousov Zhabotinsky (BZ) reaction media, toppling dominoes, plasmodium of physarum polycephalum, and the propagation of flames in a forest fire. Excitable media are characterised by the property that waves of disturbance can propagate through a medium without diminishing. The physical phenomenon that gives rise to this behaviour differs from one medium to another. The toppling domino  and the forest fire substrates are examples of single-use media --- a disturbance can propagate through a region once only. For the sake of simplicity I consider only structured excitable media, where excitations are confined to predefined regions, and do not consider unstructured free-space systems, where excitations propagate and interact in a homogeneous medium \cite{toth2009}.

One way of using excitable media for information processing is to exploit the property that some media have whereby two travelling excitations close to each other can exert an influence over each other. The consequences of that influence can be detected by exploiting another property: relative to their direction of movement, excitations may spread more favourably in some directions than in others. Flames propagating in a forest fire tend to spread out in all directions --- a flame front passing by a region of unburnt forest will not leave that part unburned. In contrast, a toppling domino will only topple in a direction perpendicular to its base. A toppling domino will not cause its neighbour to topple unless that neighbour happens to stand in the way of the toppling domino. In between these two extremes we find other types of propagation behaviour which can favour one locus over another. Planar BZ-reaction media can be made to exhibit directionally dependent propagation in a number of ways \cite{motoike1999,gorecki2009}. The preference  of plasmodium of physarum polycephalum for one direction over another can be influenced by many different factors \cite{adamatzky2010}. In certain circumstances the behaviour of propagating plasmodium has striking similarities to the behaviour of propagating waves in BZ reaction media  \cite{adamatzky2007,adamatzky2008} and can similarly be made to exhibit directionally dependent propagation (Chapter 7 of \cite{adamatzky2010}).

A recent survey of some of the ways in which structured BZ-reaction media have been used for information processing is given by G\'orecki \cite{gorecki2009}. Excitable media have been simulated at a low level of abstraction using partial differential equation models, and these low-level models have been used (in conjunction with experiments) to gain qualitative insight into the behaviour of the media and to discover relatively high-level structures such as Boolean logic gates which can be put together to make complex circuits. I attempt to fill a gap between these two levels of abstraction by quantitatively describing the time-dependent behaviour of regions of structured excitable media in response to excitations. Describing structures at this level of abstraction retains timing information, which can be exploited to make more efficient circuits, or to make systems with behaviour that cannot be succinctly described in terms of combinational or sequential logic circuits.

If a pattern of excitation is applied to an excitable medium it will evolve over time in a generally deterministic way. If another excitation is applied to the medium as it is evolving, the way in which it evolves may change, and it might unfold into a different sequence of patterns. In general, the way that the system evolves over time is dependent on the timing and location of the excitations that it receives. Within concurrency theory the term \emph{reactive system} is often used to describe a system whose behaviour depends on the pattern and timing of events that influence the system \cite{harel1985}. I would like to avoid potential confusion by making it clear at this point that the term \emph{reactive system} has nothing to do with the fact that some excitable media make use of chemical reactions.

In summary, the two problems that I address in this paper are:

\begin{enumerate}
\item How can regions of excitable media be described as \emph{reactive systems}?
\item How can the behaviour of complex structures made from excitable media be deduced from the behaviour of their constituent structures?
\end{enumerate}

This work has similar aims to some of the uses of concurrency techniques in asynchronous circuit design \cite{wang2007}. One difference is that in asynchronous design these techniques are used for the specification and verification of relatively large systems whereas here the emphasis is on \emph{exploring} and uncovering the dynamics of relatively small structures. Another difference is that in asynchronous systems few assumptions are made about timing, but here all signals are assumed to propagate at a known speed.

The purpose of this work is not simply to simulate the behaviour of structures made from excitable media at an abstract level --- this can be done perfectly well using cellular automaton models of excitable media. Rather, this work is concerned with describing the complete set of behaviours that a structure can exhibit, so that these behaviours can be understood and reasoned about without having to run exhaustive simulations of a system.

\section{Fork structures in excitable media \label{channelled-structures}}

\begin{figure}
  \begin{minipage}[t]{0.5\linewidth}
    \centering
    \includegraphics[width=1.7in]{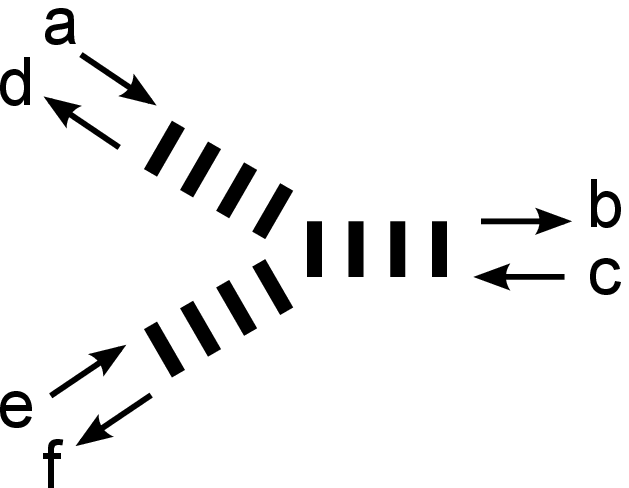}
    \caption{A domino fork.\label{domino_fork}}
  \end{minipage}
  \begin{minipage}[t]{0.5\linewidth}
    \centering
    \raisebox{13mm}{\includegraphics[width=1.7in]{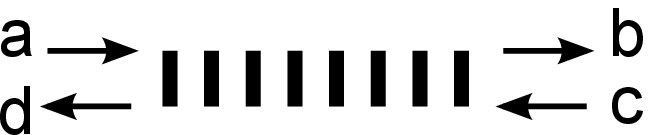}}
    \caption{A domino line.\label{domino_line}}
  \end{minipage}
\end{figure}

Figure \ref{domino_fork} shows a domino fork. This is a useful structure to use for the exposition which follows because it is a simple, familiar system. It is arguably the simplest structure in an excitable medium that can be used as the sole basis for more complex information processing structures such as Boolean logic gates and circuits.

The behaviour of the domino fork is summarised in the following paragraphs. Here the word `event' is used to mean the toppling of a domino. Dominoes can topple in any one of two directions. An `input event' to a system of dominoes is a domino outside the system toppling towards the system. An `output event' is a domino inside the system toppling away from and out of the system. The word `signal' is used to mean a travelling wave of excitation.

The domino fork has three regions on its boundary in which events can occur. In each region there are two possible events that can occur: an input event, and an output event. We use a different label for each event. Clearly in a system of dominoes once one event has happened, the other event cannot happen. In reusable excitable media, which can propagate excitations in the same region again and again, this is not the case. The definitions of `input event' and `output event' given above permit an input event and output event in the same region to occur at the same time: it is possible for a domino outside the system to topple towards the system at the same instant that a domino inide the system topples away from the system. The two dominoes will meet each other in the middle of toppling.

If event $c$ occurs then events $d$ and $f$ will occur a short time later, unless events $a$ or $e$ occur in the meantime.

If event $a$ or event $e$ occur then event $b$ will occur a short time later, unless event $c$ occurs in the meantime.

If we ignore one of the two branches of a fork, we end up with a simple line of dominoes, shown in Figure \ref{domino_line}, where input event $a$ will cause output event $b$ a short time later, unless event $c$ occurs in the meantime (with identical behaviour in the other direction). 

O'Keefe \cite{okeefe2009} showed that, subject to certain timing constraints and assuming that bridges can be built to cross one line of dominoes over another, the domino fork can be used to build Boolean circuits.
I showed that any Boolean circuit can be built, that bridges are not required, and that an asynchronous scheme can be used in which there are no timing constraints \cite{stevens2011b}, albeit with a larger number of forks for a given circuit than in O'Keefe's scheme.

\begin{figure}
\centering
\includegraphics[width=1in]{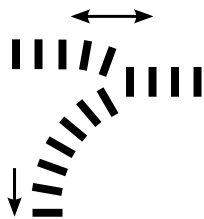}
\caption{The domino fork as a direction detecting gate.\label{direction_detecting}}
\end{figure}

A domino fork can be used as a direction-detecting gate. To help conceptualise this, imagine bending the arms of the fork into a T shape (Figure \ref{direction_detecting}), and regard the horizontal channel as a channel in which a signal may propagate in either direction, while the vertical channel is treated as an output. Now a signal travelling from right to left in the horizontal channel will cause an output to emerge from the vertical channel, but a signal travelling from left to right will not result in an output from the vertical channel.

The behaviour of a domino line can be regarded from the perspective of collision-based computing \cite{adamatzky2002}. Here, a domino line represents a channel into which objects --- in this case toppling wavefronts --- can be sent. If two objects are sent into the channel in opposite directions they will collide and annihilate with each other, but if only one object is sent in it will emerge from the other end of the line. Therefore a line effects an AND-NOT operation: if event $a$ AND-NOT event $c$ then event $b$ (subject to timing constraints). Domino fork structures placed at either end of a line can be seen as transducers which control the injection of objects into the line, and which catch objects emerging from the line.

Both the AND-NOT behaviour of a domino line and the direction-detecting behaviour of a fork seem to be essential for being able to construct more complex information processing structures using forks connected by lines. 

\begin{figure}
  \begin{minipage}[t]{0.5\linewidth}
    \centering
    \includegraphics[width=2in]{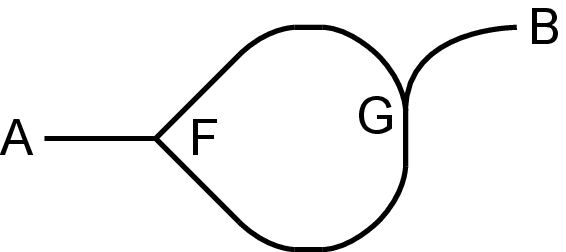}
    \caption{A one way line.\label{domino-onewayline}}
  \end{minipage}
  \begin{minipage}[t]{0.5\linewidth}
    \centering
    \includegraphics[width=1in]{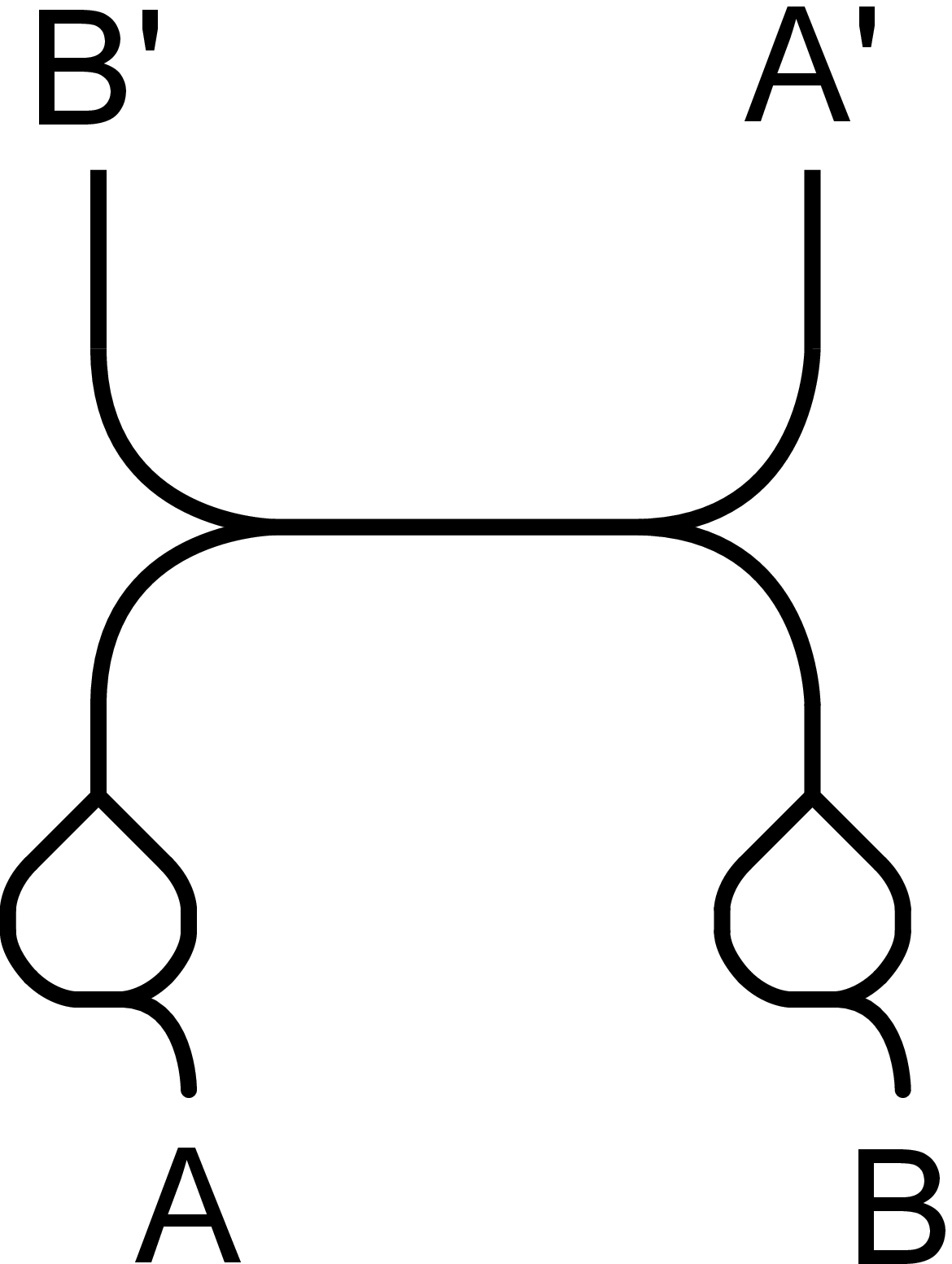}
    \caption{A single-line crossover.\label{domino-singlelinecrossover}}
  \end{minipage}
\end{figure}

In \cite{stevens2011b}, several simple structures made from domino forks and lines were described, before it was shown how they can be put together to make arbitrary Boolean circuits. Two of these simple structures are shown in Figures \ref{domino-onewayline} and \ref{domino-singlelinecrossover}.

Figure \ref{domino-onewayline} shows a one way line constructed using two forks. A
signal entering the configuration from A will split at fork F into two signals
that will collide with each other, preventing any signal from emerging at B. In
the other direction, a signal entering the configuration from B will pass through
fork G, through F and then emerge from A.

Figure \ref{domino-singlelinecrossover} shows a single-line crossover (so-called because it allows two single lines to cross over, whereas another structure in \cite{stevens2011b} allows two pairs of lines to cross over). It has two inputs A and B and two outputs
A' and B' which are crossed over topologically. A signal may arrive either at
input A or at input B (but not both). A signal arriving at A will propagate to A', but not to B or B'. A signal arriving at B will propagate to B', but not to A or A'.

\begin{figure}
  \begin{minipage}[t]{0.45\linewidth}
    \centering
    \includegraphics[width=1.5in]{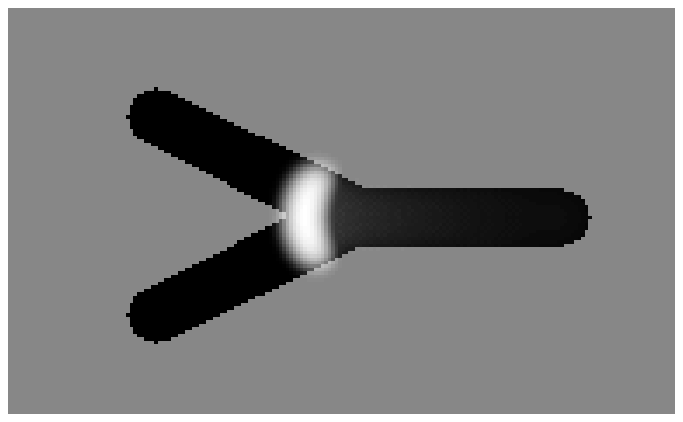}
    \caption{Splitting at a fork junction in sub-excitable BZ-reaction media.\label{bzfork1}}
  \end{minipage}
  \hspace{6mm}
  \begin{minipage}[t]{0.45\linewidth}
    \centering
    \includegraphics[width=1.5in]{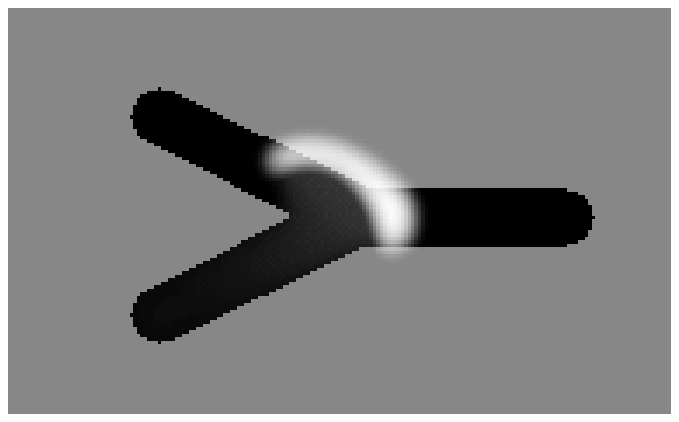}
    \caption{Directional propagation at a fork junction in sub-excitable BZ-reaction media.\label{bzfork2}}
  \end{minipage}
\end{figure}

The fork structure can also be implemented in other excitable media.  Motoike and Yoshikawa give an example of a fork-like structure in simulated BZ-reaction media \cite{motoike1999} and use it for implementing an OR gate by making use of its behaviour for the cases when $a$ or $e$ or both receive excitations (with reference to Figure \ref{domino_fork}). Motoike and Yoshikawa used unexcitable barriers to prevent backward propagation from one fork input to another, and as a consequence their fork effectively has no $c$ input. An alternative way to implement a fork structure is to use a light sensitive medium, in which the excitability of the medium can be controlled by the level of illumination \cite{gaspar1983}. Within a certain range of illumination, the medium can be made to exhibit sub-excitable behaviour, in which excitations in channels cannot propagate around sharp corners. Figures \ref{bzfork1} and \ref{bzfork2} show two snapshots from a simulation of a fork structure in this medium. Here the dark regions support sub-excitable waves, and the lighter surrounding regions inhibit excitation. This behaviour has also been observed experimentally in \cite{costello2011}.

The behaviour of the fork structure in sub-excitable BZ-reaction media is very similar to the behaviour of the domino fork, and depends on the directional propagation of excitations in the sub-excitable regime. The most significant difference between BZ-reaction media and domino media is that a BZ-reaction medium is reusable: once the medium has recovered from an excitation it can be used again. For some structures this can lead to a significantly different range of behaviours than those exhibited by a single use medium. This is discussed in section \ref{reusable-regions}.

\begin{figure}
\centering
\includegraphics[width=3in]{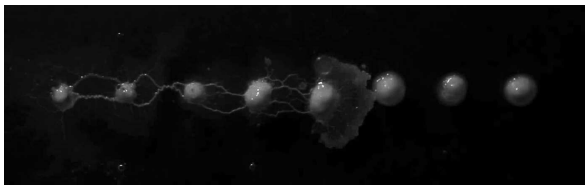}
\caption{The spreading and contracting behaviour of plasmodium of physarum polycephalum.\label{smc-actual}}
\end{figure}

\begin{figure}
  \begin{minipage}[t]{0.5\linewidth}
    \centering
    \includegraphics[width=1.5in]{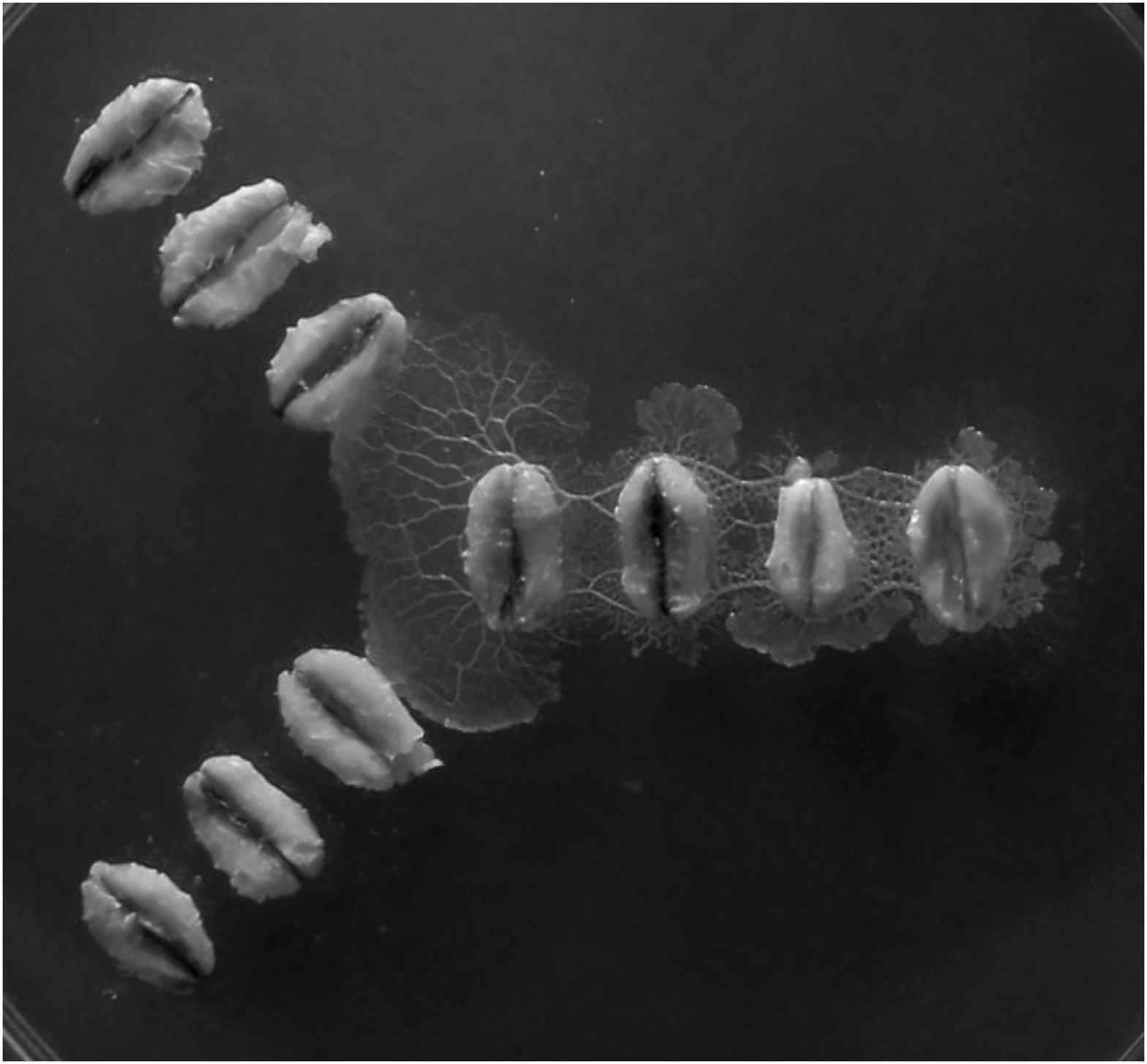}
    \caption{Physarum exhibiting a splitting behaviour at a fork junction.\label{smc-fork1}}
  \end{minipage}
  \hspace{6mm}
  \begin{minipage}[t]{0.5\linewidth}
    \centering
    \includegraphics[width=1.5in]{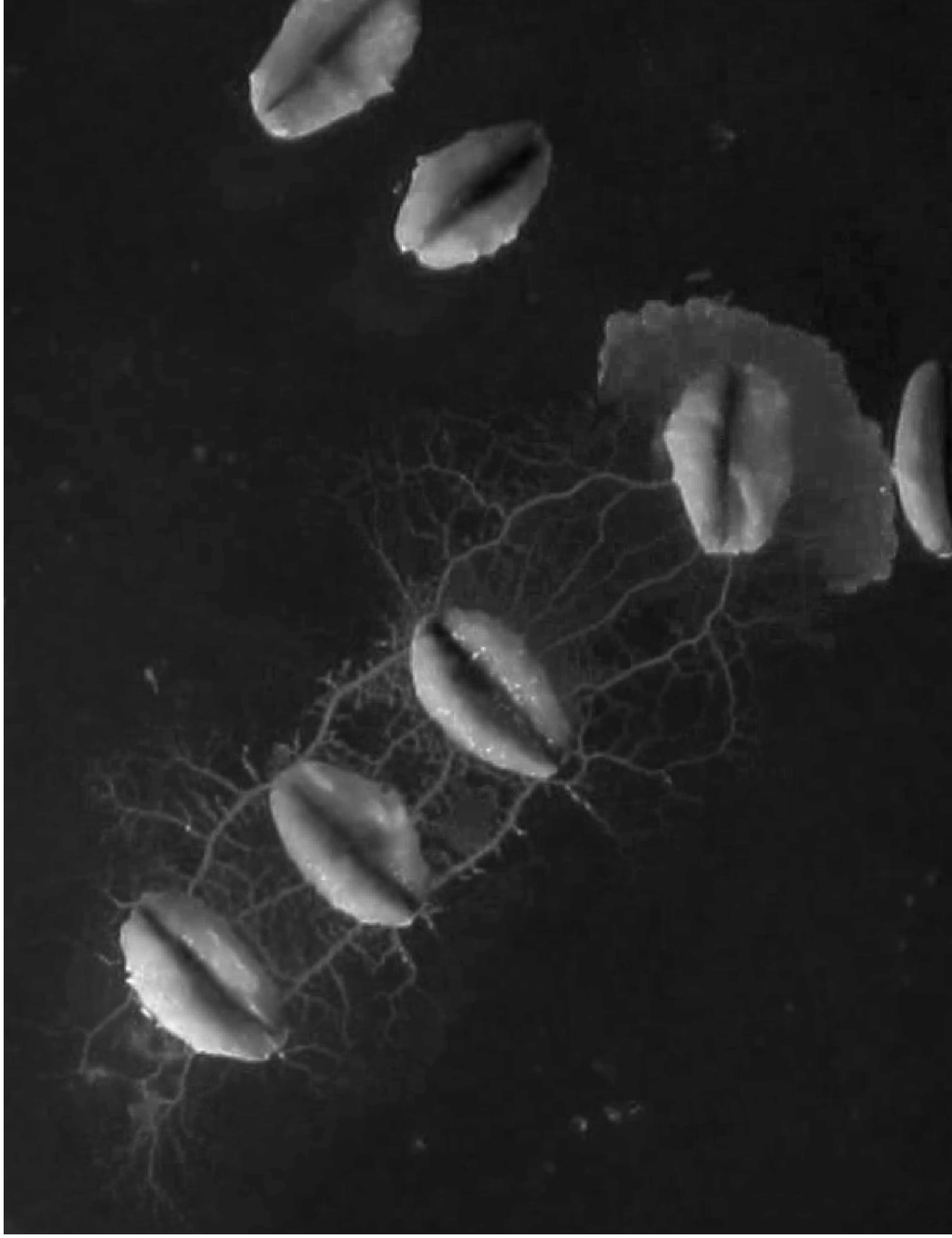}
    \caption{Physarum exhibiting direction dependent propagation behaviour, induced by the food geometry.\label{smc-fork2}}
  \end{minipage}
\end{figure}

Plasmodium of physarum polycephalum has a mode of behaviour in which it moves along a regularly spaced line of food particles in a spreading and contracting mode \cite{halvorsrud1998}. If food particles are placed close together along a line then a plasmodium will spread from a food particle until the front of the spreading area encounters another food particle. It will quickly slow and stop its spread, then migrate to the new food particle, leaving a tube connecting with the old particle. It will then begin spreading again from the new particle, repeating the same behaviour. Figure \ref{smc-actual} shows an image of a plasmodium exhibiting this behaviour. Figures \ref{smc-fork1} and \ref{smc-fork2} show work in progress to coax physarum to exhibit fork-like behaviour, obtained from the spreading and contracting mode of propagation with a carefully designed food geometry.

Fork structures and line structures are used exclusively in the exposition that follows. The method that is introduced should work equally well for other structures with behaviour that can be described in terms of a reactive system.

\section{Transition systems}

I begin by defining a type of transition system that is capable of modelling the behaviour of single fork and line structures in terms of input and output events, and the timing relationships between them. I then give a set of rules for composing transition systems that will allow us to deduce the behaviour of larger and more complex structures from the behaviour of fork and line structures. When the total range of behaviours that a complex structure can exhibit is difficult to comprehend, I show how to select portions of its behaviour which can be understood in isolation and which can help to understand the total behaviour.

The transition systems defined here make use of the description of transition systems given in \cite{winskel1995}. The meaning given to transitions is similar to that used in SCCS \cite{milner1989} in so far as every transition represents the passage of a unit of time, in addition to the occurrence of zero or more events, but instead of using event sets as labels, Boolean expressions over event symbols are used. A common clock is assumed among all transition systems, so transitions in any pair of transition systems occur together at the same time.

Notation and concepts will be introduced by giving examples along with more formal descriptions. Whenever the term `transition system' is used from here onwards, it refers to the particular type of transition system developed here, rather than to transition systems in general. 

\begin{definition}
A transition system is a tuple $(S,i,L,Tran,E,I)$ where $S$ is a set of states with initial state $i$, $L$ is a set of labels, $Tran \subseteq S \times L \times S$ is the transition relation, $E$ is a set of event symbols, and $I \subseteq E$ is a set of input event symbols. Symbols in $E \setminus I$ are output event symbols.
\end{definition}

If we are referring to two or more different transitions systems then in order to distinguish one from the other we use a dot notation to refer to the elements of a tuple: If $P$ is a transition system then $P.S$ is the set of states of $P$, $P.i$ is the initial state of P, etc.

Event symbols represent the occurrence of an event in the physical system that the transition system represents. For the sake of brevity we will use the word `events' to refer to both event symbols and the events that the symbols represent. 

The set $I$, which we will call the input interface, is a set of events that are generated outside the transition system and which the system reacts to. The set $E \setminus I$, which we will call the output interface, is a set of events that are generated by the system. 

A transition system represents a structure made from excitable media. Each state corresponds to a particular pattern of excitation, and transitions from one state to another correspond to the evolution over time of patterns of excitation, possibly subject to the effects of input events. The input interface of the transition system corresponds to locations where excitations can be fed into the structure, and the output interface corresponds to locations where excitations can emerge from the structure for observation or to feed into the inputs of other structures.

Labels are Boolean expressions over events, where the symbol $a$ in an expression corresponds to the occurrence of event $a$, and the expression $\neg a$ corresponds to the non-occurrence of event $a$. We will use $Bool(A)$ to refer to the set of Boolean expressions over events from the set $A$. In our Boolean expressions we use $\cdot$ for conjunction, $+$ for disjunction and $\neg$ for negation. Two labels are equivalent if they represent the same Boolean function, regardless of how that function is expressed. A label represents a collection of sets of events. We choose to use Boolean expressions (rather than event sets as in SCCS) for labels because by doing so we often reduce the number of transitions that are needed in a transition system, because we can specify many sets of events in a single transition, rather than having a separate transition for each set. The synchronisation operator for composing two transition system remains simple --- it is essentially the conjunction of two Boolean expressions.

\begin{definition}
We say that a label expression $l$ with event variables $a_j$ \emph{matches} a set of input events $A \subseteq I$ if, given the truth assignment $a_j = True$ \textrm{ if } $a_j \in A, False \textrm{ if } a_j \in I \setminus A$, the resulting expression  is not equal to $False$. i.e. it is either equal to $True$, or it has some remaining variables not in $I$.
\end{definition}

We must place some additional constraints on transition systems to bring their behaviour into closer agreement with the physical systems we are interested in. We define these constraints here, and later on when we compose two transition systems we will need to show that these constraints still hold in the transition system that is produced as the result of the composition.

A transition corresponds to the passage of a unit of time in addition to the occurrence of events. We can think of the occurrence of events as being interleaved with the passage of units of time. Events occur, then a unit of time elapses, then more events occur, and so on. When a particular set of input events occurs, the transition followed is that which matches the events that occur. A transition system must be deterministic: no set of input events should match more than one transition from a given state. Additionally, we will adopt the convention that there must always be a transition that matches any given set of input events. Even if a system reaches a state from which it can never leave, we require that it has a transition back to the same state, matching any set of input events. The reason for this convention is that it simplifies the rule for composing two systems.

\begin{constraint}
For each $(s,A) \in S \times \mathcal{P}(I)$ there must be exactly one $(s,l,s') \in Tran$ such that $l$ matches $A$.

\label{label-satisfiable}
\end{constraint}

A label must be the conjunction of two subexpressions, one being any Boolean expression of input events, the other being a conjunction of output events or their negation, in which every output event appears. This ensures that every transition effectively specifies which output events occur during that transition --- they are the events that are not negated in the output event subexpression.

We first define a set of expressions that are conjunctions of variables or their negation:

\begin{definition}
Let $Conj(A) = \{ b_1 \cdot b_2 \cdot ... \cdot b_n \mid (b_1,b_2,...,b_n) \in \prod \{a_j,\bar{a_j}\}\}$ for $a_j \in A$
\end{definition}

\begin{constraint}
$L = \{x \cdot y \mid x \in Bool(I) \textrm{ and } y \in Conj(E \setminus I) \}$
\label{label-form}
\end{constraint}
 
One of the advantages of using Boolean expressions representing collections of event sets for labels is that if a particular input event has absolutely no effect on the behaviour of a transition system in a given state --- that is, the destination state is the same regardless of whether that input event occurs or not --- then we can omit that event from any label in transitions from that state. A transition system in a state where some inputs have no effect could correspond to a physical system in a state where an input feeds into a region of excitable medium which has not yet recovered from a previous event. More generally it corresponds to a physical system evolving along a trajectory on which a particular input makes no difference to that trajectory.

The rule for determining which transition a transition system will undergo in response to a set of input events, and which output events it will produce is given by the following definition:

\begin{definition}
If a system is in state $s$ and the set of input events that occurs is $A$
then the next state $s_{next}$ of the system is given by the element $(s,l,s_{next}) \in Tran$
for which $l$ matches $A$. The set of output events that the system will produce are those which appear non-negated in the output event subexpression of $l$.
\end{definition}

\begin{definition}
A state $s$ is reachable if and only if it is the initial state $i$, or there is a transition $(s',l,s) \in Tran$ where $l$ is satisfiable and $s'$ is reachable.
\end{definition}

\begin{constraint}
All states in a transition system must be reachable.
\label{state-reachable}
\end{constraint}

Finally, the next constraint specifies that no input can have an instantaneous effect on any output. There must be a delay of at least one time unit between the occurrence of a set of events and a change in the output behaviour of the system. This is realised by making sure that all labels for transitions from a given state have the same output event subexpression.

\begin{constraint}
For each $s \in S$ there is a single expression $y_s \in Conj(E \setminus I)$  such that for every $(s,l,s') \in Tran$, $l$ is of the form $x \cdot y_s$ (where $x \in Bool(I)$ is different for each transition).
\label{single-output-expression}
\end{constraint}

It is reasonable to insist that any transition system that corresponds to a physical excitable media system satisfies constraint \ref{single-output-expression}. If we were to permit a transition system to flout constraint \ref{single-output-expression}, then we could end up in a contradictory situation where two systems that shared both inputs and outputs could specify, say, that events $a$ and $b$ must always occur simultaneously, and also that $b$ can only occur when $a$ does not occur.

These definitions and constraints permit us to specify transition systems that represent all of the possible behaviours of a fork and a line. Before doing so we will consider some simpler examples and then define rules for composing two transition systems. These simple examples do not correspond to any structures in excitable media. They serve simply to illustrate the concepts that have been described so far and introduce in an incremental way the notation and conventions used for representing transition systems graphically.

\begin{figure}
\centering
\includegraphics[width=2.6in]{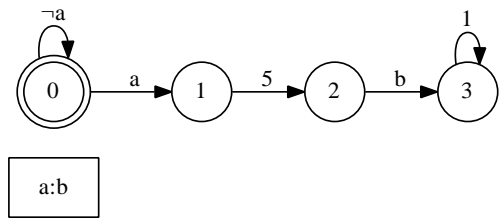}
\caption{An example of a transition system.\label{graphexample1}}
\end{figure}

Figure \ref{graphexample1} is an example of a simple transition system. The input interface for this system is ${a}$ and the output interface is ${b}$. This is displayed as $a:b$ in the box in the bottom left of Figure \ref{graphexample1}. The initial state is represented using one circle inside another and is also named as state 0. Other states are given numeric names. Transitions are represented as arrows from one state to another, with the label for the transition written above the arrow.

This example introduces two conventions for representing transition systems that make them easier to read. The first convention is the way that label expressions are written. Labels could be written simply as Boolean expressions, but this representation cannot easily be read at a glance, so instead we take account of constraint \ref{label-form} regarding the allowable sub-expression for output events, and write the sub-expression corresponding to the input event variables first, followed by a colon, followed by a set of those output event variables that are not negated in the label expression. For example, if $a$ and $b$ are input events, and $c$ and $d$ are output events then the label expression $(a+e)\bar{c}d$ is written $(a+e):d$.  The colon is omitted if the output event subexpression is a conjunction of negated output event variables, or if the label expression matches any set of input events. If the label expression is simply a conjunction of negated output events then we write the label as $1$.

The second convention concerns the graphical representation of a consecutive sequence of transitions from one state to another that is not influenced by any input events and which does not produce any output events. Since such a sequence of states and transitions corresponds to the passage of a length of time equal to the length of the sequence, we write it using a single arrow with a number above the arrow representing the length of the sequence. In Figure \ref{graphexample1}, the arrow with the number 5 written above it between states 1 and 2 has the same meaning as if we drew four extra states and transitions between state 1 and 2, and labelled each of the five transition between states 1 and 2 with the label 1.

The behaviour of the transition system in Figure \ref{graphexample1} can be summarised as follows. If the system is in its initial state and event $a$ does not occur then it will remain in the initial state no matter how many time units elapse. If the system is in its initial state and input event $a$ occurs then the system will be in state 1 after 1 unit of time has elapsed. After a further 5 units of time have elapsed the system will enter state 2. It will then generate output event $b$ and will be in state 3 after 1 unit of time. Once the system is in state 3 it will remain in state 3 regardless of how much time elapses.

\begin{figure}
\centering
\includegraphics[width=2.2in]{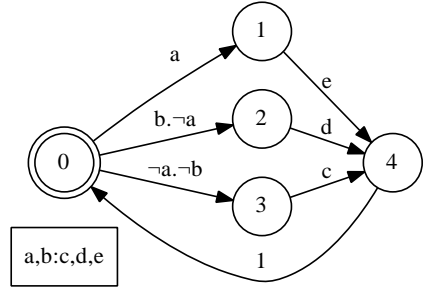}
\caption{A more complex example of a transition system.\label{graphexample2}}
\end{figure}

Figure \ref{graphexample2} shows a slightly more complex example. The input interface of this system is ${a,b}$ and the output interface is ${c,d,e}$. If the system is in its initial state and input event $a$ occurs, or if both $a$ and $b$ occur, it will enter state 1 after 1 unit of time. It will then generate event $e$ and enter state 4 after a further unit of time. If the system is in its initial state and input event $b$ occurs by itself, it will enter state 2 after 1 unit of time. It will then generate event $d$ and enter state 4. If the system is in its initial state and neither input event occurs then one unit of time will elapse and the system will enter state 3. It will then output event $c$, and enter state 4 after 1 unit of time. From state 4, the system will return to the initial state after 1 unit of time.

\subsection{Composing transition systems}
 
\begin{figure}
\centering
\includegraphics[width=1.18in]{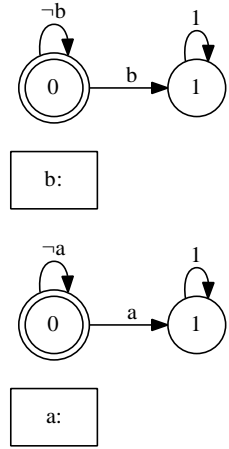}
\caption{Two simple transition systems.\label{simplebeforecomposing}}
\end{figure}

\begin{figure}
\centering
\includegraphics[width=2in]{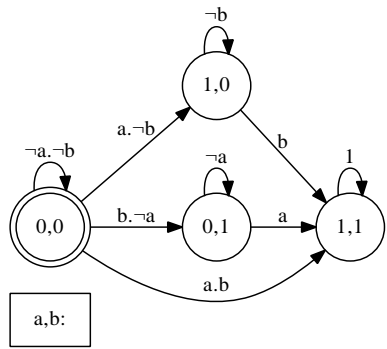}
\caption{The composition of the two independent systems from Figure \ref{simplebeforecomposing}.\label{simpleaftercomposing}}
\end{figure}

In general, the behaviour of two transition systems $P$ and $Q$ operating independently and concurrently can be described by a single transition system by considering each state in $P$ in combination with each state in $Q$. Figure \ref{simplebeforecomposing} shows two simple transition systems that have no events in common. Although these two transition systems operate independently, both are in their initial state at the starting time and they operate under a common clock. Because both of the transition systems in Figure \ref{simplebeforecomposing} have  transitions returning to every state, each of the systems can remain in the initial state or the final state for any length of time while the other system makes a state transition. Figure \ref{simpleaftercomposing} shows the result of composing these transition systems.

The notation $P \parallel Q$ will be used for the composition of $P$ and $Q$. When $P$ and $Q$ are composed the interface of $P$ may have events in common with the interface of $Q$, so we must come up with a set of rules for combining transitions to make sure that the occurrence or non-occurrence of events in $P$ is synchronised with the occurrence or non-occurrence of events in $Q$.

For two transition systems to be composable, they must both satisfy constraints \ref{label-satisfiable}, \ref{label-form} and \ref{state-reachable}, and any event that they have in common must be an input in one system and an output in the other.

Regarding constraint \ref{single-output-expression}, we will allow that either $P$ or $Q$ does not satisfy this constraint. The reason for this is made clear later on in section \ref{reusable-regions}. However, whichever of $P$ or $Q$ does not satisify constraint \ref{single-output-expression} must have no inputs coming from the other system, and all of its outputs must feed into the other system.

\begin{definition}
Two transition systems $P$ and $Q$ are composable if and only if the following conditions hold:

\begin{itemize}
\item[\bf{A}] Both $P$ and $Q$ satisfy constraints \ref{label-satisfiable}, \ref{label-form} and \ref{state-reachable}.

\item[\bf{B}] If $P$ does not satisfy constraint \ref{single-output-expression}, then $Q$ must satisfy constraint \ref{single-output-expression} and $P.E \cap Q.E = P.E \setminus P.I$.

\item[\bf{C}] If $Q$ does not satisfy constraint \ref{single-output-expression}, then $P$ must satisfy constraint \ref{single-output-expression} and $P.E \cap Q.E = Q.E \setminus Q.I$.

\item[\bf{D}] $a \in P.E \cap Q.E \implies (a \in P.I \textrm{ and } a \notin Q.I) \textrm{ or } (a \in Q.I \textrm{ and } a \notin P.I)$.

\end{itemize}

\label{composability} 
\end{definition}

To compose two transition systems $P$ and $Q$ to produce a transition system $T = P \parallel Q$ we first form the product $T'$, where labels in transitions are combined by conjunction:

\begin{definition}
The product $T' = P \times Q$ is defined as:

$T'.S = P.S \times Q.S$

$T'.i = (P.i,Q.i)$

$T'.Tran = \{ ((s_P,s_Q),l_P \cdot l_Q,(s'_P,s'_Q)) \mid (s_P,l_P,s'_P) \in P.Tran$ and $(s_Q,l_Q,s'_Q) \in Q.Tran \}$

$T'.L = \{ l \mid (s,l,s') \in T'.Tran \}$

$T'.E = P.E \cup Q.E$

$T'.I = P.I \cup Q.I$
\end{definition}

We obtain $T$ by discarding states in $T'.S$ that are not reachable and discarding transitions in $T'.Tran$ that are not satisfiable. We also remove any reference to events that appear in both $P.E$ and $Q.E$ --- these are events that correspond to locations where two regions of excitable media meet, so they do not appear in the interface of $P \parallel Q$.

\begin{definition}
For $l_P \in P.L$, $l_Q \in Q.L$, where $l_P \cdot l_Q$ is satisfiable, let $\theta(l_P \cdot l_Q)$ be the expression obtained by setting any event variable $a \in P.E \cap Q.E$ that appears in $l_P \cdot l_Q$ to the value that leaves the resulting expression satisfiable. Constraint \ref{label-form} and definition \ref{composability}D ensure that there will only be one such value for each $a$. 
\label{theta-definition}
\end{definition}

\begin{definition}
The composition $T = P \parallel Q$ is defined as:

$T.S = \{ s \in T'.S \mid s \textrm{ is reachable in } T' \}$

$T.i = T'.i$

$T.Tran = \{(s,\theta(l_P \cdot l_Q),s') \mid (s,l_P \cdot l_Q,s') \in T'.Tran \textrm{ and } l_P \cdot l_Q \textrm{ is satisfiable}\}$

$T.L = \{ k \mid (s,k,s') \in T.Tran \}$

$T.E = T'.E \setminus (P.E \cap Q.E)$

$T.I = T'.I \setminus (P.E \cap Q.E)$
\label{composition}
\end{definition}

We will now show that $T = P \parallel Q$ satisfies constraints \ref{label-satisfiable}, \ref{label-form}, \ref{state-reachable} and \ref{single-output-expression}.

\begin{proposition}
$T$ satisfies constraint \ref{label-satisfiable}.
\end{proposition}

\begin{proof}
Take any $((s_P,s_Q),A) \in T.S \times \mathcal{P}(T.I)$.

If $P$ satisfies constraint \ref{single-output-expression} then let $y_P$ be the output subexpression corresponding to $s_P$. Otherwise, by definition \ref{composability}B, $Q.E \cap P.I = \emptyset$  (i.e. no outputs from $Q$ are inputs to $P$), and constraint \ref{label-satisfiable} applied to $P$ implies that there is exactly one $(s_P,l_P,s'_P)$ such that $l_P$ matches $A \cap P.I$, and let $y_p$ be the output subexpression of $l_P$.

Similarly, if $Q$ satisfies constraint \ref{single-output-expression} then let $y_Q$ be the output subexpression corresponding to $s_Q$. Otherwise, $P.E \cap Q.I = \emptyset$ and there is exactly one $(s_Q,l_Q,s'_Q)$ such that $l_Q$ matches $A \cap Q.I$, and let $y_Q$ be the output subexpression of $l_Q$. 

Between them, the two subexpressions $y_P$ and $y_Q$ determine a single $C \in \mathcal{P} (P.E \cap Q.E)$ for a given $((s_P,s_Q),A) \in T.S \times \mathcal{P}(T.I)$. Constraint \ref{label-satisfiable} applied to $P$ and $Q$ implies that there is exactly one $(s_P,l_P,s'_P)$ such that $l_P$ matches $A \cup C \cap P.I$ and exactly one $(s_Q,l_Q,s'_Q)$ such that $l_Q$ matches $A \cup C \cap Q.I$.

Therefore, for a given $(s_P,s_Q)$ there must be exactly one transition with label $\theta(l_P \cdot l_Q)$ that matches $A$. Thus constraint \ref{label-satisfiable} holds for $P \parallel Q$.
\end{proof}

\begin{proposition}
$T$ satisfies constraint \ref{label-form}.
\end{proposition}

\begin{proof}
$l$ is of the form $v \cdot w \cdot x \cdot y$ where $v \in Bool(P.I), w \in Conj(P.E \setminus P.I), x \in Bool(Q.I)$ and $y \in Conj(Q.E \setminus Q.I)$.

$\theta(l)$ is obtained from $l$ by setting all variables in $P.E \cap Q.E$ to $True$ or $False$, so $\theta(l)$ is of the form $v' \cdot w' \cdot x' \cdot y'$ where $v' \in Bool(P.I \setminus (P.E \cap Q.E)), w' \in Conj(P.E \setminus (P.I \cup P.E \cap Q.E)), x' \in Bool(Q.I \setminus (P.E \cap Q.E)), y' \in Conj(Q.E \setminus (Q.I \cup P.E \cap Q.E)))$. This can be written as $(v' \cdot x') \cdot (w' \cdot y')$. Now $v' \cdot x' \in Bool(T.I)$ and $w' \cdot y' \in Conj(T.E \setminus T.I)$, so constraint \ref{label-form} is satisfied.
\end{proof}

\begin{proposition}
$T$ satisfies constraint \ref{state-reachable}.
\end{proposition}

\begin{proof}
$T$ was obtained it from $T'$ by discarding those states that were not reachable. 
\end{proof}

\begin{proposition}
$T$ satisfies constraint \ref{single-output-expression}.
\end{proposition}

\begin{proof}
Since $P$ and $Q$ both satisfy constraint \ref{label-form}, for a given state $(s_P,s_Q) \in T.S$, the label of every transition
$((s_P,s_Q),\theta(l_P \cdot l_Q),(s_P',s_Q'))$ can be written as
$\theta(x_P \cdot y_P \cdot x_Q \cdot y_Q) = \theta(x_P \cdot x_Q) \cdot \theta(y_P) \cdot \theta(y_Q)$.

If $P$ satisfies constraint \ref{single-output-expression} then $y_P$ is determined soley by $s_P$ (i.e. it is independent of $s_P'$), and so is $\theta(y_P)$. Otherwise, by definition \ref{composability}B, all variables in $y_P$ are in $P.E \cap Q.E$, and so $\theta(y_P) = True$, so again $\theta(y_P)$ is determined by $s_P$.
By the same argument, $\theta(y_Q)$ is determined by $s_Q$.

Therefore $\theta(y_P) \cdot \theta(y_Q) \in Conj(T.E \setminus T.I)$, is a subexpression common to all transitions beginning from $(s_P,s_Q) \in T.S$, so $T$ satisfies constraint \ref{single-output-expression}.
\end{proof}

\section{Examples \label{examples}}

\subsection{Describing and composing fork and line structures}

We begin by considering single-use fork and line structures. Systems made from single-use structures behave like toppling domino systems. Each event can occur only once, and the system cannot return back to its initial state. In section \ref{reusable-regions} we will deal with regions that recover after a period of time and which can be used again. 

The transition systems given below for fork and line structures should be treated as axiomatic. No justification is offered here for why these transition systems describe the behaviour of fork and line structures: it is assumed that fork and line structures are sufficiently simple that the correspondence between their behaviour and the transition systems given below is obvious. However, once we have specified transition systems for the fork and the line, the behaviour of any structure built from these can be deduced automatically from the behaviour of the fork and the line by using the composition rules given above. 

\begin{figure}
\centering
\includegraphics[width=1.3in]{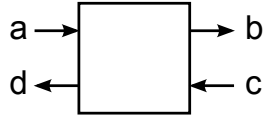}
\caption{A unit length horizontal channel in an excitable medium.\label{excitableregion}}
\end{figure}

\begin{figure}
\centering
\includegraphics[width=2in]{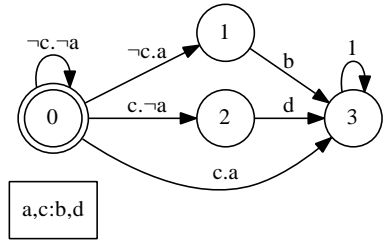}
\caption{A transition system that describes the behaviour of the region in Figure \ref{excitableregion}.\label{excitableregiongraph}}
\end{figure}

Figure \ref{excitableregion} shows a unit length region of a horizontal channel in an excitable medium. The region has two input events $a$ and $c$, and two output events $b$ and $d$. Event $a$ will cause event $b$, event $c$ will cause event $d$, but if both input events occur at the same time, there will be no output events. Figure \ref{excitableregiongraph} shows a transition system that describes this behaviour.

\begin{figure}
\centering
\includegraphics[width=1.6in]{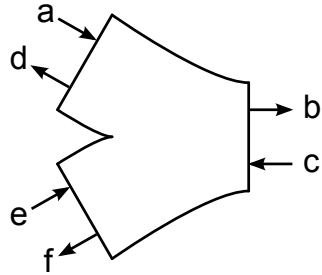}
\caption{A unit length fork in an excitable medium.\label{excitablefork}}
\end{figure}

\begin{figure}
\centering
\includegraphics[width=2.275in]{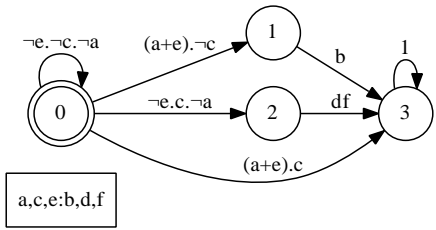}
\caption{A transition system that describes the behaviour of the fork in Figure \ref{excitablefork}.\label{excitableforkgraph}}
\end{figure}

Figure \ref{excitablefork} shows a unit length fork structure in the medium. Figure \ref{excitableforkgraph} shows a transition system that describes the behaviour of the fork, based on the informal description given in section \ref{channelled-structures}.

\begin{figure}
\centering
\includegraphics[width=2.1in]{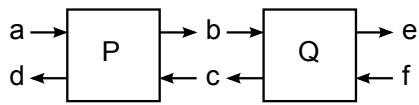}
\caption{Two composed unit length horizontal channels.\label{excitableregioncompose}}
\end{figure}

To compose two of the excitable regions shown in Figure \ref{excitableregion}, let $P$ be one region and let $Q$ be a neighbouring region. Because we will be composing $P$ and $Q$ we would like event $b$ of $P$ to correspond to one of the input events of $Q$, and event $c$ of $P$ to correspond to one of the output events of $Q$, according to Figure \ref{excitableregioncompose}. We use the notation $Q = P [ m ]$ to mean that Q is a transition system exactly like P, but with events renamed according to the mapping $m$. In this case we use the mapping $m = \{ (a,b),(d,c),(b,e),(c,f) \}$

\begin{figure}
\centering
\includegraphics[width=2.5in]{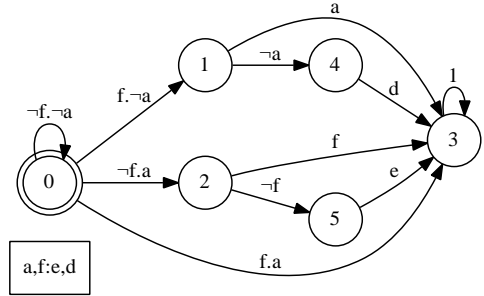}
\caption{The transition system that describes the composition shown in Figure \ref{excitableregioncompose}.\label{twocomposedregions}}
\end{figure}

By definition \ref{composition}, we can deduce $P \parallel Q$ from $P$ and $Q$ to obtain the transition system shown in Figure \ref{twocomposedregions}. This transition system matches our intuitive understanding of the behaviour of $P \parallel Q$: event $a$ will cause event $e$ two time units later, unless event $f$ occurs before the effect of $a$ has time to propagate to $e$. Similarly, $f$ will cause $d$ unless it is interrupted by $a$. We can make a line of any length by composing unit length regions in this way.

\begin{figure}
\centering
\includegraphics[width=3.0in]{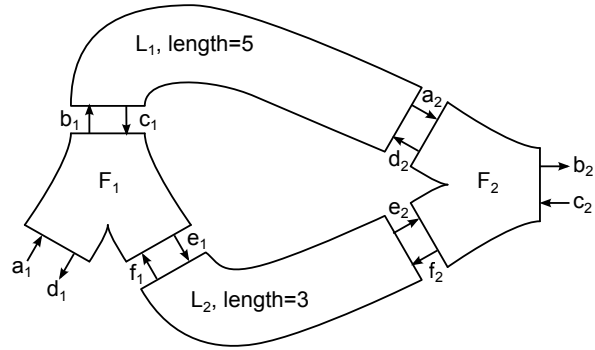}
\caption{A one way line can be decomposed into four simpler components.\label{onewaycomponents}}
\end{figure}

Next we consider the single-use one way line shown in Figure \ref{domino-onewayline}. Figure \ref{onewaycomponents} shows how this can be constructed from two forks $F_1$ and $F_2$, a line $L_1$ of length 5 and a line $L_2$ of length 3.
$L_1$ and $L_2$ are made by composing unit length regions, with their ends relabelled to match the events of $F_1$ and $F_2$ that they correspond to.

\begin{figure}
\centering
\includegraphics[width=4.0in]{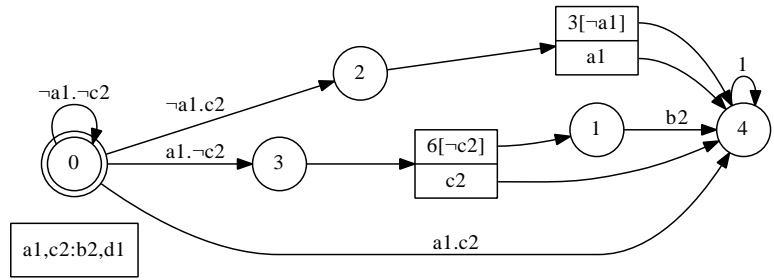}
\caption{A transition system describing the behaviour of the one-way line $F_1 \parallel L_1 \parallel L_2 \parallel F_2$.\label{onewaytransitionsystem}}
\end{figure}

\begin{figure}
\centering
\includegraphics[width=3in]{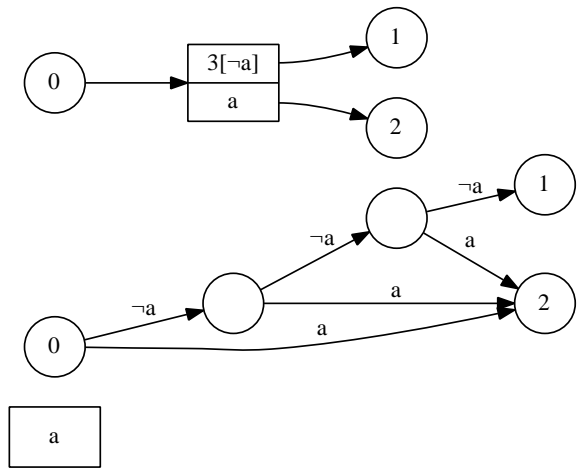}
\caption{The interrupt box notation above is equivalent to the system of states below.\label{interruptequivalence}}
\end{figure}

Figure \ref{onewaytransitionsystem} shows the result of the composition $F_1 \parallel L_1 \parallel L_2 \parallel F_2$. This figure introduces another notational convenience. Just as we use an arrow with an integer larger than 1 written above it to denote a series of transitions of duration 1, so the split boxes in Figure \ref{onewaytransitionsystem} represent an interruptible sequence: a series of transitions that are followed so long as events that occur satisfy the expression inside the square brackets in the upper part of the box, but which can be interrupted on the occurrence of events that satisfy the expression in the lower part of the box. The number outside the square brackets in the upper part of the box is the length of the sequence. The arrow leading from the upper part of the box leads to the state that will be reached if the interrupting events do not occur, that leading from the lower part leads to the state that will be reached if they do. To illustrate this further, Figure \ref{interruptequivalence} depicts an interrupt box and the collection of states and transitions that it denotes. Sequences of states that fit patterns similar to this occur frequently in the type of medium considered here --- for example an excitation propagating along a line of length $n$ will reach the end after $n$ units of time, unless interrupted by an excitation coming in the other direction.

It can be seen from Figure \ref{onewaytransitionsystem} that the occurrence of $a_1$ will lead to the occurrence of $b_2$ seven time units later (unless interrupted by $c_2$), but the occurrence of $c_2$ does not lead to any occurrence of $d_1$.

\subsection{Reusable regions \label{reusable-regions}}

\begin{figure}
\centering
\includegraphics[width=2in]{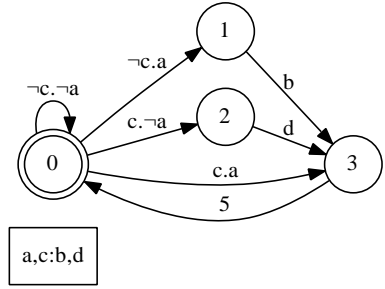}
\caption{A transition system for a unit length reusable region of an excitable medium.\label{excitableregionreusegraph}}
\end{figure}

So far we have considered single-use regions. We can easily change a transition system describing a single-use unit length fork or line region into a transition system for a region that can be re-used after a `recovery time' by adding a delay transitions from the final state back to the initial state. Figure \ref{excitableregionreusegraph} shows a transition system for the reusable equivalent of Figure \ref{excitableregion} with a recovery time of 5 time units. The choice of five time units is somewhat arbitrary, but leads to behaviour that is qualitatively similar to that obtained in BZ reaction media, where a propagating wave leaves behind it a short-lived inhibiting tail. The inhibiting tail prevents the propagation of further waves until it has disappeared.

Let us consider a reusable one-way line, with the same structure as that in Figure \ref{onewaycomponents}, but where each unit length region is reusable with a recovery time of 5 time units. The transition system for this reusable one-way line has 41 states. The reason for such a large number of states is that as an event propagates through the system, different regions of the structure will recover at different times, and at any time after the inputs to the one-way line have recovered, further input events can occur.

We can manage this extra complexity to some extent by placing constraints on the way in which a structure may be used. The context in which we are using a structure may provide us with knowledge about the number of times each input event can occur, or the rate at which they may occur. For example, we can specify that each input event must occur exactly once, in any order, with an arbitrarily long period of time separating the two events. This will not necessarily result in the same collection of behaviours as the single-use one way line, because we still allow the possibility that regions within the structure can be re-used.

In order to specify this we can construct a second transition system $R$ that restricts the pattern of input events to those that we are interested in, and then form $T \parallel R$ to produce a transition system limited to the behaviour determined by $R$.

Let $(F_1 \parallel L_1 \parallel L_2 \parallel F_2)[\{(a_1,a_1'),(c_2,c_2')\}]$ be the transition system for a reusable version of the one way line from Figure \ref{onewaycomponents}, with $a_1$ renamed to $a_1'$, and $c_2$ renamed to $c_2'$. 

\begin{figure}
\centering
\includegraphics[width=3in]{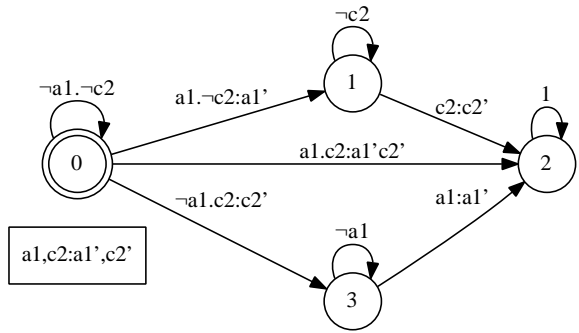}
\caption{A restriction transition system in which $a_1'$ and $c_2'$ both occur exactly once.\label{restriction-system}}
\end{figure}

\begin{figure}
\centering
\includegraphics[width=5.5in]{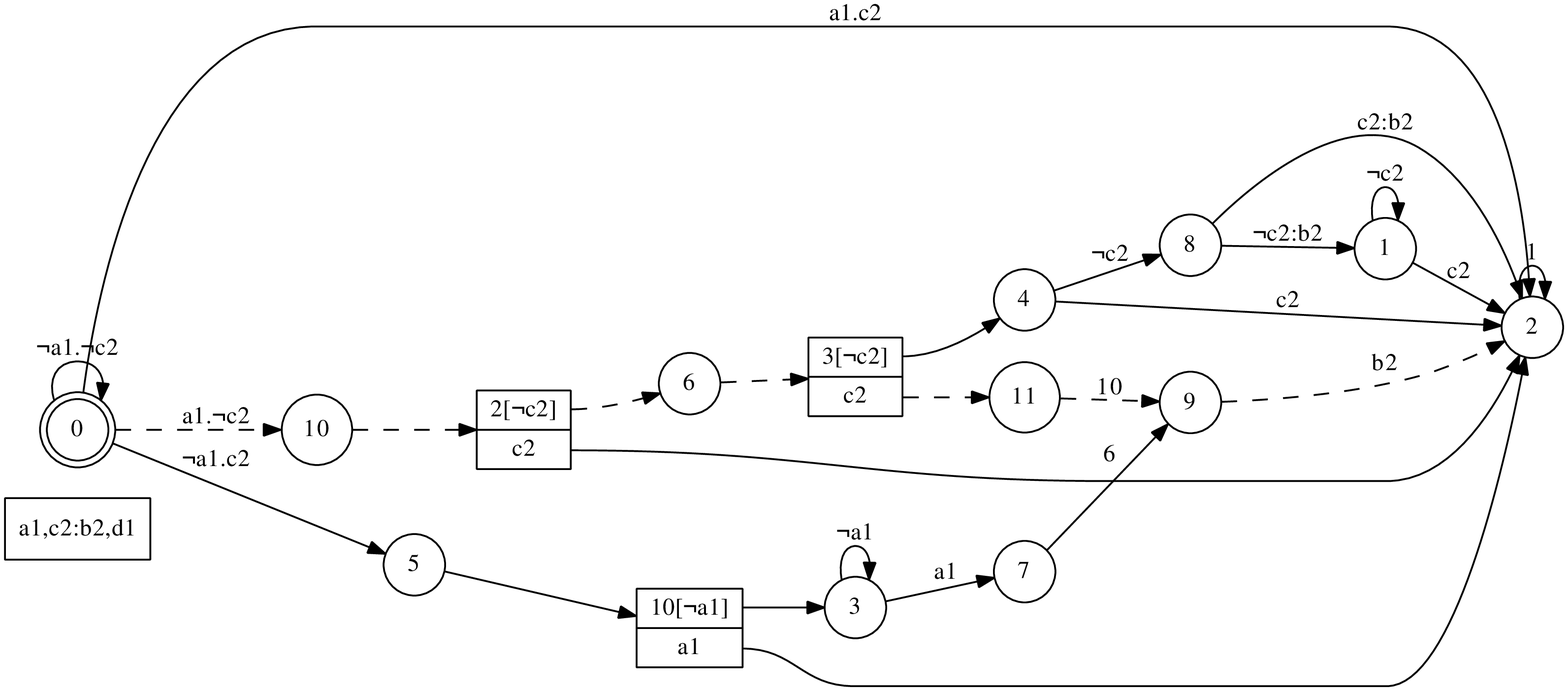}
\caption{A transition system for the reusable one way line in which events $a_1$ and $c_2$ occur exactly once.\label{onewayreuse}}
\end{figure}

Figure \ref{restriction-system} shows a transition system $R$ with input events $\{a_1,c_2\}$ and output events $\{a_1',c_2'\}$ in which the first occurrence of $a_1$ immediately leads to $a_1'$, and the first occurrence of  $c_2$ immediately leads to $c_2'$. Each of the output events $a_1'$ and $c_2'$ occurs exactly once in any path from state 0 to state 2 in Figure \ref{restriction-system}. $R$ does not satisfy constraint \ref{single-output-expression}: outputs $a_1'$ and $c_2'$ of $R$ mirror the corresponding inputs immediately without any delay (for the first occurrence of each input). However, because $R$ has no inputs connected to outputs of the system it is being composed with, and because all outputs of $R$ feed into the system it is being composed with, the composition $((F_1 \parallel L_1 \parallel L_2 \parallel F_2)[\{(a_1,a_1'),(c_2,c_2')\}]) \parallel R$ does satisfy constraint \ref{single-output-expression}. This composition is shown in Figure \ref{onewayreuse}.

The reason why the rules for composability given in definition \ref{composability} are formulated to allow one of the transition systems being composed to flout constraint \ref{single-output-expression} is to permit us to do what we are doing here: they permit us to compose a transition system with a restriction transition system in which certain output events mirror certain input events.

\begin{figure}
\centering
\includegraphics[width=5in]{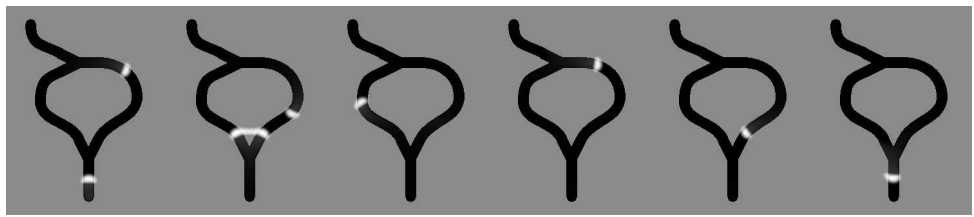}
\caption{A one way line exhibiting the behaviour corresponding to the sequence of dashed arrows in Figure \ref{onewayreuse}.\label{oneway_delaybehaviour}}
\end{figure}

Figure \ref{onewayreuse} has several things in common with the single-use one-way line described by Figure \ref{onewaytransitionsystem} --- in each case $a_1$ will cause event $b_2$ if not interrupted by $c_2$, but in Figure \ref{onewaytransitionsystem} the occurrence of event $c_2$ before event $b_2$ is due to occur will always prevent $b_2$ from occurring. In Figure \ref{onewayreuse} however, there is a window of time during which the occurrence of $c_2$ after $a_1$ will delay the occurrence of $b_2$ rather than prevent it. The transitions for this behaviour are shown as dashed arrows in Figure \ref{onewayreuse}. This is a behaviour of the reusable one-way line that the single-use version cannot exhibit. It arises from the possibility that regions within the structure can be excited more than once, even though the input events may not occur more than once. Figure \ref{oneway_delaybehaviour} shows successive snapshots of a simulation of a sub-excitable BZ-reaction medium exhibiting this behaviour.

\subsection{A modulo-2 counter \label{modulo-2 counter}}

\begin{figure}
\centering
\includegraphics[width=2.8in]{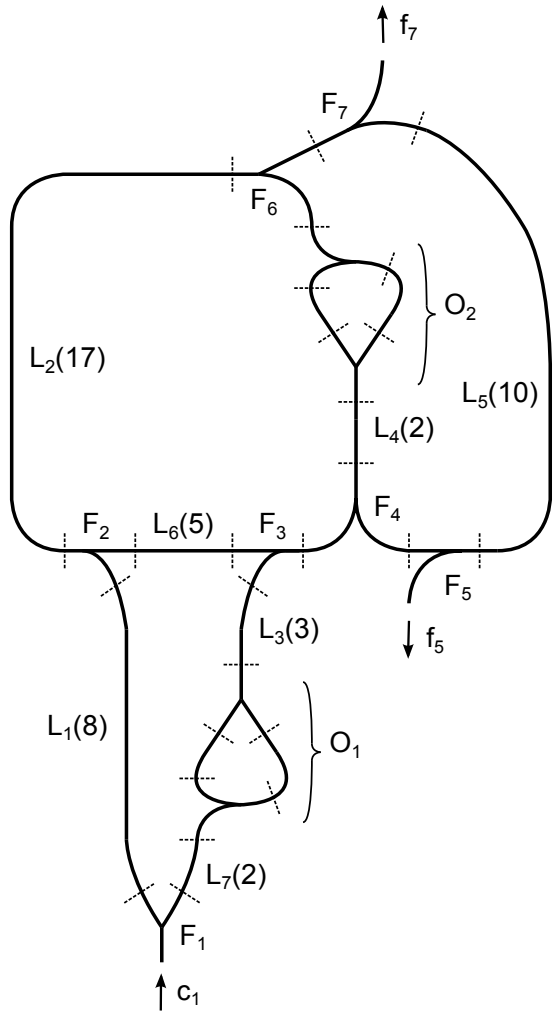}
\caption{A structure that can behave like a modulo-2 counter.\label{loopexample}}
\end{figure}

The final example considered here is a modulo-2 counter, shown in Figure \ref{loopexample}. This example is made from  reusable excitable regions. It contains two one-way lines labelled $O_1$ and $O_2$, seven forks labelled $F_1$ to $F_7$ and seven lines labelled $L_1$ to $L_7$. The length of each line is written in brackets in Figure \ref{loopexample}. This example contains a loop structure made from $L_2, F_6, O_2, L_4, F_4, F_3, L_6$ and $F_2$ around which an excitation can propagate endlessly. The event $c_1$ will lead to an excitation in this loop as follows: an excitation from $c_1$ will split at fork $F_1$, one excitation will propagate along $L_1$ and into the loop travelling clockwise, the other will propagate through $O_1$ and into the loop travelling anti-clockwise. The latter will reach $O_2$ and stop well before the former reaches $O_2$ in the opposite direction, so the former will propagate endlessly around the loop. 

\begin{figure}
\centering
\includegraphics[width=3in]{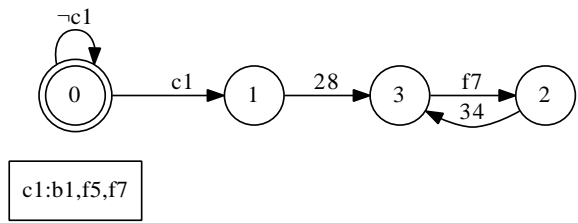}
\caption{A transition system for Figure \ref{loopexample} when only one input excitation is applied.\label{loopsingle}}
\end{figure}

This is confirmed by the transition system in Figure \ref{loopsingle}, which is produced from the parallel composition of all of the components, after specifying that no excitation will ever be input into the forks from which events $f_5$ and $f_7$ emerge, and that only one excitation will be fed into $c_1$. The fork $F_7$ provides an output $f_7$ which can be used to `see' the excitation every time it propagates around the loop.

Every time around the loop, the excitation will split at $F_6$ and one excitation will travel along $L_5$. It will not reach $f_5$ because it will be annihilated before it reaches $F_5$ by an excitation splitting from the loop at $F_4$. 

Once there is an excitation in the loop, it will periodically progress along $L_3$ into $O_1$. It will be stopped by $O_1$. If a second input is applied to $c_1$ at the right moment, it will coincide with the first excitation in the loop, so will not cause a second clockwise-travelling excitation. It will also meet the first excitation in $O_1$. Its progress through $O_1$ will effectively be delayed, according to the behaviour of $O_1$ given by the dashed line in Figure \ref{onewayreuse}, so it will arrive at $O_2$ late enough to cause the first excitation to be delayed in its progress around the loop, also according to the behaviour in Figure \ref{onewayreuse}. This will allow the excitation travelling along $L_5$ to reach output $f_5$, and also to reach the loop and annihilate the excitation in the loop.

To summarise: a single excitation at $c_1$ will cause an excitation to propagate around the loop. A second excitation at $c_1$ (at the right moment) will lead to an output at $f_5$, and will return the system back to its original state. Therefore an output will emerge from $f_5$ for every other event at $c_1$ (subject to timing constraints). This is the behaviour of a modulo-2 counter.

\begin{figure}
\centering
\includegraphics[width=5in]{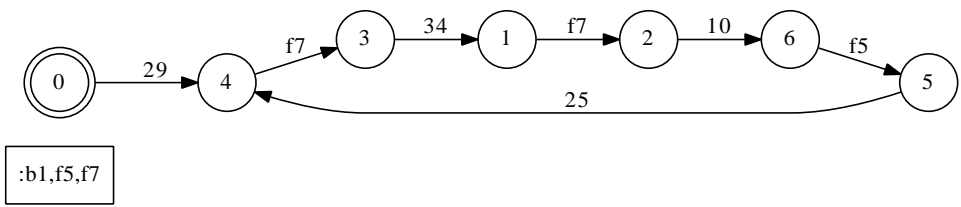}
\caption{A transition system for Figure \ref{loopexample} when an input excitation is applied every 36 time units.\label{loopdouble}}
\end{figure}

Inspection of the 191-state transition system (not shown here) for the case when $c_1$ occurs twice reveals that in order for the system to exhibit this behaviour, the interval between the first and second occurrences of $c_1$ must be 35,36 or 37 time units. By connecting a transition system that produces an output every 36 time units to input $c_1$, the transition system shown in Figure \ref{loopdouble} is obtained. This shows that, as expected, there is one occurrence of $f_5$ for every two occurrences of $f_7$, and that the system reverts to a previous state after the occurrence of $f_5$.

\section{Implementation}

An implementation of a set of routines for manipulating transition systems, written using the Python 3 programming language, is available from:

\vspace{6mm}
\url{http://www.srm.org.uk/downloads/transition.py}
\vspace{6mm}

This source file contains functions for creating all of the examples that are used in this paper. The Python programming language was used because it has built in data types for representing sets, tuples and maps, because it has a concise syntax, and because it is freely available and widely supported. The source file contains comments which will help with understanding the implementation.

\section{Discussion and conclusions}

This paper has demonstrated that it is possible to use transition systems to describe the behaviour of unit length fork and line structures in excitable media, and that it is possible to deduce the behaviour of larger structures from the behaviour of their components. Although this paper has exclusively used fork and line structures for making systems, it should be possible to use the method given here for automatically deducing the behaviour of other excitable media structures made from simple components, where the behaviour of the simple components can be described by transition systems.

I have chosen to model behaviour that involves the propagation and interaction of excitations in channelled structures, but there are other types of behaviour that excitable media can exhibit, and which can be used for information processing, that the methods used in this paper cannot describe. The use of a planar BZ-reaction medium for image processing \cite{kuhnert1989} exploits a large rectangular section of the medium and so cannot be described in terms of excitations propagating along channels.
Steinbock, Kettunen and Showalter \cite{steinbock1996} make use of the interaction of several waves over an extended area to implement logic gates. A transition system could be used to describe one of these logic gates and its timing properties, but it could not be used to describe the lower-level extended-area interactions of excitations and deduce the behaviour of a logic gate from these.

Much research has been carried out on the use of process algebra to describe and reason about concurrent systems. There are two main reasons why I chose to work directly with transition systems and use a graph-based representation for a set of behaviours, rather than make use of process algebra and adopt a linguistic representation for a set of behaviours. Firstly, process algebraic notation can often obscure the simplicity that is readily apparent in a graphical representation: even for small systems, not all sets of behaviours can be represented in an easy-to-understand way in any of the current linguistic representations employed in process algebra. Secondly, the range of meanings and structures that can be given to labels in transition systems provides enough flexibility to permit the physical systems considered here to be described, but the structure of transition systems is constrained enough to reduce the danger of losing the ability to compose systems together. 

However, we have only considered relatively simple systems in this paper and I do not claim to have fully explored all of the possible ways of using concurrency theory to model structured excitable media. For more complex systems a well-formulated algebraic approach that permits equational reasoning and recursive definitions may be useful. The sequence and interrupt box notations in Figures \ref{graphexample1} and \ref{onewaytransitionsystem} respectively were introduced as syntactic conveniences for representing collections of states and transitions. Instead of regarding these simply as syntactic entities they could alternatively be treated as semantic entities, and rules for manipulating them during composition could perhaps be formulated, without having to expand them into collections of transitions and states.

Several simplifying assumptions and approximations have been made. I have assumed that all regions of the medium that we are modelling propagate excitations at the same rate. This assumption is valid for simulated planar BZ-reaction media. It is also reasonable in experimental planar BZ-reaction media, where experimental conditions can easily be achieved in which the propagation of a wave of excitation through two identically structured regions takes a very similar length of time. Theoretical and experimental investigations of domino lines indicate that their behaviour is also predictable and repeatable \cite{wagon2005}. This assumption is less valid for plasmodium of physarum polycephalum, where experiments to measure the speed of propagation of plasmodia in the spreading/contracting mode \cite{halvorsrud1998} show that the speed of propagation can vary over the course of a single experiment, and can vary even more from one experimental run to another. It is not yet known how easily these variations can be reduced by controlling the experimental conditions.

I have used discrete time rather than continuous time. The effect of this is to restrict lengths in structures to integer multiples of the length that an excitation covers in a single time unit, and to restrict the times at which input events can occur to integer multiples of a single time unit. Since we are free to choose which units of physical time and length to use, this does not seem to be a severe restriction. It is unlikely that it would be possible to make a deterministic physical structure so critically dependent on timing for its operation that it cannot be modelled using integral channel lengths for some choice of length unit and time unit.

Modelling a physical system using a transition system involves identifying the states of the physical system with states of the transition system, and identifying changes of state in the physical system with transitions in the transition system. Trajectories of a system correspond to sequences of states and transitions. For synchronous systems where each trajectory has a common clock, an alternative approach can be conceived of in which trajectories rather than states are treated as primary, and transitions in a system are from one trajectory to another rather than from one state to another. If such a scheme can be formalised, and if parallel composition rules for such a trajectory-oriented scheme can be formulated, then this might be a more natural way of modelling the type of system considered in this paper.
 
\section{Acknowledgements}

The work was partially supported by Leverhulme Trust grant F/00577/1. Thanks to Andrew Adamatzky for useful comments on an earlier draft of this paper.

\FloatBarrier

\end{document}